\documentclass{article}

\usepackage{graphicx}
\usepackage[compress]{cite}

\usepackage{amsmath}   
\usepackage{amssymb}   
		       
\def\eq#1{{Eq.~(\ref{#1})}}

\def\cc{{cosmological\ constant}}

\def\md{microscopic degrees of freedom of the spacetime}

\def\dqg{density of states of the quantum spacetime}

\title{The atoms of spacetime and the cosmological constant}

\author{Thanu Padmanabhan,\\
IUCAA, Pune University Campus,\\
Ganeshkhind, Pune, India.\\
email: paddy@iucaa.in}

\date{ }

\begin{document}

\maketitle

\begin{abstract}
I describe an approach which relates  classical gravity to the quantum microstructure of spacetime. In this approach, the field equations arise from maximizing the density of states of the matter plus geometry. The former is identified using the thermodynamics of null surfaces.  The latter arises from the existence of a zero-point length in the spacetime which associates 
 an internal degree of freedom with each event in the spacetime, in the form of a fluctuating vector of constant norm. The density of states, as well as the 
resulting field equations, remain invariant 
under the shift $T^a_b\to T^a_b+$(constant)$\delta^a_b$ (arising from the addition of a constant to the 
 matter Lagrangian, which is a symmetry of the matter sector). The \cc\ ($\Lambda$) arises as an integration constant and renders the amount of cosmic information $(I_c)$ accessible to an eternal observer finite. The relation between $\Lambda$ and $I_c$ allows us to determine the numerical value of $(\Lambda)$ from the information content of the quantum  spacetime, within the context of cosmology.\footnote{This is an updated version of the lecture I gave at the DICE 2016 meeting. I will concentrate here on the conceptual aspects of the approach. More technical details can be found in Ref.\cite{ijmpdreview2016,lwf}.}
\end{abstract}

\section{Gravity from the atoms of space: Summary}
\label{sec:grfratoms}

It is possible to obtain the gravitational field equations from an extremum principle  based on the density of states of the quantum geometry. I will motivate a suitable definition of the \dqg\ ---  which is the key new idea in this approach --- that leads to deeper insights into the \md\ and,  as a bonus, provides \cite{ijmpdreview2016}  a fresh perspective on the numerical value of the \cc.  In this section, I will summarize the results
and highlight the conceptual aspects.\footnote{The signature is $(-,+,+,+)$ and I use natural units with $c=1, \hbar =1$ and set $\kappa = 8\pi G =8\pi L_P^2$ where $L_P$ is the Planck length $(G\hbar/c^3)^{1/2}=G^{1/2}$ in natural units.  Latin letters $i, j$ etc. range over spacetime indices and the Greek letters $\alpha, \beta$ etc. range over the spatial indices. I will write $x$ for $x^i$, suppressing the index, when no confusion is likely to arise.} Later sections have some mathematical details and I refer the reader to two recent reviews \cite{ijmpdreview2016,lwf} for a more extensive discussion.

 Let me begin by recalling   that the physics of the fluids can be presented at two different levels, one more fundamental than the other. The first level is the continuum description which ignores the fact that there are microscopic degrees of freedom in the fluid in the form of atoms/molecules. This approach uses  variables like density $\rho(x^i)$, pressure $P(x^i)$, temperature $T(x^i)$, fluid momentum $P^\mu(x^i)$, etc. and the dynamics  is governed by the continuity equation and, say, the Navier-Stokes equation.    Of course, the existence of temperature (proportional to the random kinetic energy of the atoms) as well as the  transport coefficients tells us that such a description is fundamentally inconsistent. The second, deeper, level of description uses  the  distribution function $f(x^i, p_j)$, which \textit{counts}  the number of  atoms $dN = f(x^i, p_j)d^3xd^3p$   per unit phase space volume $d^3xd^3p$ (with the constraint $p^2=m^2$ making the phase space  six-dimensional).
 This description, in terms of a distribution function, is remarkable because it allows us to use the continuum language and --- at the same time --- recognize the discrete nature of the fluid. (One could equivalently think of $f$ as the number of \textit{degrees of freedom} per unit phase space volume.)
 The new feature which arises at this level of description is the  ``internal'' variable $p^\mu$ which allows us to  describe atoms with different \textit{microscopic} momenta co-existing at the same event $x^i$. The \textit{macroscopic} momentum  of the fluid $P^\mu(x^i)$ is given by the average value $\langle p^\mu\rangle =P^\mu(x^i)$. Similarly, we have 
 $\langle p^\mu p^\nu\rangle =P^\mu(x^i)P^\nu(x^i)+\Sigma^{\mu\nu}$ where the second term arises from the dispersion of momentum.
 
 In a similar manner, I want to introduce a function $\rho_g (x^i,\phi_A)$ to describe the \dqg. Here, $\phi_A$ (with $A=1,2,3,...$) denotes possible internal degrees of freedom (analogous to the momentum $p_i$ for the distribution function for the molecules of a fluid) which  exist as fluctuating \textit{internal} variables at \textit{each} event $x^i$. Their behaviour,  at any event $x^i$, is determined by a probability distribution $P(\phi_A,x^i)$, the form of which depends on the microscopic quantum state of the spacetime. Of course, we cannot determine this function without knowing more about the quantum structure of spacetime, but --- fortunately, as we will see --- we only need some properties of the  average values like $\langle \phi_A\rangle, \langle \phi_A\phi_B\rangle$ etc., for the purpose of obtaining and interpreting the classical field equations of gravity.
 The dependence of $\rho_g (x^i,\phi_A)$ on $x^i$ arises only indirectly through   the geometrical variables like the metric tensor, curvature tensor etc., (which I will collectively denote as $\mathcal{G}_N(x)$ with $N=1,2,3,...$), so that $\rho_g (x,\phi_A)=\rho_g (\mathcal{G}_N(x), \phi_A)$. Such a description is not exact but will be valid at some mesoscopic scales larger than $L_P$ so that the variables  $\mathcal{G}_N(x)$ can be defined. At these scales, the Planck length plays a role  which is analogous to that of the mean-free-path in the kinetic theory.

 It turns out that there is a natural way of defining $\rho_g (x^i,\phi_A)$, if we introduce discreteness into the spacetime through a zero-point length. Remarkably enough, this procedure \textit{also}  identifies  the internal variable $\phi_A$ as a constant norm four-vector $n^a$,  which can be thought of  as a microscopic, fluctuating, quantum variable at each event $x^i$. In terms of this internal, vector degree of freedom, the $\rho_g(x^i,n_a)$ is given by
\begin{equation}
\ln \rho_g \propto \left[1- \frac{L_P^2}{8\pi} R_{ab}(x) n^an^b\right] 
\label{rhogresult}
\end{equation}
 The fluctuations of $n^a$ are governed by some  probability functional $P[n^i(x),x]$, which is the probability that the quantum geometry is described by a vector field  $n_a(x)$ at every $x$. As I said before the form of $P$  is unknown at present; but, fortunately, we will need only two properties of this  probability distribution $P[n^i(x),x]$ which can be derived: (i) It preserves the norm of $n^i$, which is unity in the Euclidean sector and  zero  in the Lorentzian sector; i.e  $P[n^i(x),x]$ will have the form $F[n(x),x]\delta(n^2-\epsilon)$ with $\epsilon=1$ in the Euclidean space and zero in the Lorentzian spacetime. (ii) The   average of $n^a$ over the fluctuations (in a given quantum state of the geometry) gives,\footnote{This is analogous to the average value of the microscopic momenta of fluid particles $\langle p^\mu\rangle =P^\mu(x^i)$ giving rise to the macroscopic momentum of the fluid. The key difference is that, in normal fluid mechanics, the distribution function itself is used to do the averaging while here the fluctuations are governed by some other probability distribution $P(x,n)$. Recall that the null normal $\ell_a$ also defines the tangent vector to the null geodesic congruence on the null surface; in this sense, it is indeed the momentum of the photons traveling along the null geodesics.} in the Lorentzian spacetime, a null normal $\ell^a(x^i)$ to a patch of null surface; i.e.,  $\langle n^a\rangle = \ell^a(x^i)$. Such  averages can be defined through the functional integral
\begin{equation}
 \langle n_a\rangle=\int\mathcal{D}n\; n_a(x)P[n^a(x),x]=\ell_a(x)
\end{equation} 
where $P[n^a(x),x]$ is  parametrized by some null vector field $\ell_a(x)$. (For example, $P$ could be a sharply peaked Gaussian functional in the variable $[n^a(x)-\ell^a(x)]$.) 
  Of course, different quantum states of the spacetime geometry will lead to  different $P$ with different null normals $\ell_a (x^i)$ as their mean values; so, in fact,  the expectation value $\langle n_a\rangle $ actually leads to the \textit{set of all null normals} $\{\ell_a(x^i)\}$ at an event $x^i$ when we take into account all quantum states. Similarly, $\langle n_in_j\rangle =\ell_i\ell_j+\sigma_{ij}$ where the second term $\sigma_{ij}$ represents quantum gravitational corrections to the mean value, etc. Therefore, the mean value $\langle \ln\rho_g(x^i,n_a)\rangle $,
 in the continuum limit, is given by:
\begin{equation}
\langle \ln\rho_g(x^i,n_a)\rangle \propto 
 1-\frac{1}{8\pi} L_{P}^{2} R_{ab}\ell^a\ell^b +....
\label{denast2}
\end{equation}
where we have not displayed terms proportional to $R_{ab}\sigma^{ab}$ which are of higher order and independent of $\ell_a(x)$.
Remarkably enough, the quantity 
\begin{equation}
\mathcal{H}_g\equiv - \frac{1}{8\pi L_P^2}R_{ab}\ell^a\ell^b
\label{defhg}
\end{equation}
which occurs here leads to a term quadratic in the \textit{derivatives}  $\nabla_a\ell_b$ (plus an ignorable total divergence) and
has an interpretation as the 
gravitational contribution to the heating rate (per unit area) of the null surface to which $\ell_a$ is the normal. Its integral over the null surface, $Q_g$, can be interpreted  as the gravitational contribution to the heat content of the null surface. 

To complete the picture, we also need the corresponding expression for the matter sector. In the continuum limit, it is straightforward to show --- using the concept of local Rindler horizons --- that  the corresponding quantity for matter is given by:
\begin{equation}
\langle \ln \rho_m\rangle  \propto  L_P^4 T_{ab} \ell^a \ell^b=L_P^4\mathcal{H}_m
\label{rhomresult}
\end{equation}
where --- as we shall see --- $\mathcal{H}_m$ can be interpreted as the heat density contributed by matter crossing a null patch.
Taking into account both matter and spacetime, the total number of degrees of freedom, in the continuum limit --- in a state characterized by the vector field $\ell_a(x)$ --- will be
\begin{equation}
 \langle \Omega_{\rm tot}\rangle_\ell  =\ \prod_{x}\, \langle \rho_g\rangle  \langle \rho_m\rangle  
 = \exp \sum_x \left( \langle \ln \rho_g\rangle  + \langle \ln \rho_m\rangle \right)\equiv \exp [S_{\rm grav}(\ell)+S_{\rm m}(\ell)]
 \label{omtot}
\end{equation}
to the lowest order (when we ignore the fluctuations, so that $\ln\langle \rho\rangle \approx\langle \ln\rho\rangle $).
The proportionality constants in \eq{rhogresult} and \eq{rhomresult} can be taken to be the same factor, say, $\mu$, by choosing the measure in this sum appropriately. 
 The  $\ell_a$ dependent part of the configurational entropy $S_{\rm tot}=S_{\rm grav}+S_{\rm m}$ is given by the functional
\begin{equation}
S_{\rm tot}[\ell(x)]=
\int_{\mathcal{S}} d^3V_x\; \mu E^a_b \langle n_an^b\rangle =\int_{\mathcal{S}} d^3V_x\;\mu\left(T^a_b (x) - \frac{1}{\kappa} R^a_b (x)\right) \ell_a(x)\ell^b(x) + ....
\label{av1}
\end{equation}
where, in the continuum limit, the sum over $x$ is replaced by integration over the null surface $\mathcal{S}$ for which $\ell^a(x)$ is the normal, with the measure
$d^3 V_x=(d\lambda d^2 x \sqrt{\gamma}/L_P^3)$ and the proportionality constant $\mu$ is introduced. 

The  gravitational field equations can be obtained by extremizing the expression for $\langle \Omega_{\rm tot}\rangle_\ell$ or, equivalently, the configurational entropy $S_{\rm tot}=S_{\rm grav}+S_{\rm m}$ over $\ell$ and demanding that the extremum condition holds for all $\ell_a$. Since different quantum states of geometry will lead to different $\ell_a(x)$ at the same event $x^i$, this is equivalent to demanding the validity of the extremum condition for all quantum states of the geometry, which are relevant in the classical limit.
The extremum of \eq{av1} with respect   to $\ell^a \to \ell^a + \delta \ell^a$, 
subject to the constraint $\ell^2=$ constant, will then lead to Einstein's equations, with a cosmological constant arising as an integration constant.\footnote{This is straightforward: Demanding that the extremum condition holds for all $\ell_a$ leads to $R^{a}_{b} - \kappa T^a_{b} = f(x) \delta^a_{b}$. Taking the divergence of this equation and using  $\nabla_a T^a_b =0$ and 
$\nabla_a R^a_b = (1/2)\partial_b R$,  you get $f(x) = (1/2)R +$ a constant, leading to Einstein's equations, with a cosmological constant arising as an integration constant.} 
Moreover, we will see later that the integrand of \eq{av1} \textit{itself} can be interpreted as the heating rate of the null surface. So, the classical limit makes perfect thermodynamic sense.

So  we obtain the field equations \textit{without} treating the metric as a dynamical variable. While it is unusual, there are strong reasons to believe that the metric tensor is probably not the correct dynamical variable to use in an extremum principle. I have described this aspect in detail in Ref. \cite{ijmpdreview2016,explore} but let me briefly highlight this historical accident, in the construction of the field equations of general relativity, because of which   an innocent geometrical object like $g_{ab}$ acquired the status of a dynamical variable in an action principle. 

Given the principle of equivalence and general covariance, you can conclude that: (a) Gravity is best described as the effect of curvature of spacetime by using a line interval $ds^2=g_{ab}dx^adx^b$. (b) The influence of gravity on other systems can be obtained by demanding the validity of special relativity in the local inertial frames and general covariance. (c) In the Newtonian limit, the only nontrivial metric coefficient is $g_{00} = - (1+2\phi)$.
What Einstein was  looking for at this stage was a generalization of the Newtonian field equation $\nabla^2 (2\phi) =  \kappa \rho$ which  can be  written as  $-\nabla^2 g_{00} = \kappa T_{00}$ where $\rho$ is identified with the time-time component $T_{00}$ of the divergence-free, second rank symmetric energy momentum tensor $T_{ab}$. 
Since $\nabla^2$ is not Lorentz invariant, you might think (alas, wrongly!) that you need to  ``generalize'' the $\nabla^2$ to $\square^2$ so that the left hand side has second derivatives with respect to \textit{both space and time} coordinates.  
The second derivatives of the metric tensor can be expressed covariantly  in terms of the curvature tensor, which led Einstein to look for a   divergence-free, second rank symmetric tensor to replace $\nabla^2 g_{00}$ in the left hand side. This leads to $G_{ab}=\kappa T_{ab}$.

But Einstein could have taken a different, and better, route! 
One can generalize the Newton's law of gravity $\nabla^2\phi\propto\rho$, \textit{retaining the right hand side as it is and without introducing second time derivatives in the left hand side}. 
To do this, note that: 
(i) The energy density  $\rho=T_{ab}u^au^b$ in the right hand side, is observer dependent where $u^i$ is the four velocity of an observer. There is no need to be afraid of  $u^i$ appearing in this equation. 
(ii) Consistency demands that we should generalize  the left hand side, $\nabla^2\phi$, to some scalar which also \textit{depends on the four-velocity $u^i$ of the observer} bilinearly. 
(iii) It is perfectly acceptable for the left hand side \textit{not} to have second \textit{time} derivatives of the metric, in the rest frame of the observer, just as they do not occur in $\nabla^2\phi$. 

To obtain such a scalar (which will replace $\nabla^2\phi$) containing  \textit{spatial} second derivatives that depend bilinearly on $u^i$,  we first project the indices of $R_{abcd}$ to the space orthogonal to $u^i$, using the projection tensor $P^i_j=\delta^i_j+u^iu_j$, thereby obtaining the tensor
$\mathcal{R}_{ijkl}\equiv P^a_iP^b_jP^c_kP^d_l R_{abcd}$. The only scalar you can construct from $\mathcal{R}_{ijkl}$ is $\mathcal{R}^{-2}\equiv\mathcal{R}_{ij}^{ij}$ where $\mathcal{R}$ can be thought of as the radius of curvature of the space.\footnote{The $\mathcal{R}_{ijkl}$ and $\mathcal{R}$ should \textit{not} to be confused with the curvature tensor $^3R_{ijkl}$ and the Ricci scalar $^3R$ of the 3-space orthogonal to $u^i$.} The natural generalization of $\nabla^2\phi\propto\rho$ is then given by $\mathcal{R}^{-2}\propto\rho=T_{ab}u^au^b$. Working out the left hand side (see e.g., p. 259 of 
Ref.~\cite{key7}), we get:
 $
G_{ab}u^au^b=\kappa T_{ab}u^au^b.                                  
$
If we demand that this relation holds for all observers, we get back $G_{ab}=\kappa T_{ab}$. A natural generalization of this approach is to demand 
$
G_{ab}\ell^a\ell^b=\kappa T_{ab}\ell^a\ell^b                                  
$
for all \textit{null} vectors $\ell_a$ which has the attractive feature that it is invariant under $T^a_b\to T^a_b+(\text{constant}) \delta^a_b$, thereby preserving a symmetry respected by matter sector. This is exactly what we get in our approach.

Even though we vary the null vector field $\ell_a$ in our extremum principle in \eq{av1}, the vector field $\ell_a$  also cannot be thought of as a dynamical variable in the conventional sense. Recall that, normally, if you vary a dynamical variable $q_i$ in an action principle, you get an evolution equation for that variable $q_i$. Here, we demand that the extremum condition should hold for all $q_i$ and use this demand to constrain the background spacetime. Such an extremum condition is more similar to thermodynamic extremum principles (describing an equilibrium configuration) rather than to action principles (leading to dynamical evolution). While a bit unusual, all this is consistent with the idea that the gravitational field equations must be interpreted in a thermodynamic language.

I will now describe how the expression for $\rho_g$ is obtained and how the internal degree of freedom $n_a$ arises.

\section{Area associated with a spacetime  event}
\label{sec:zpl}

The two primitive geometrical constructs one can think of in a spacetime  are the area and the volume. It is, therefore,  natural to assume that the \dqg, $\rho_g(\mathcal{P})$,  at an event $\mathcal{P}$, should be some function $F$ of either the area $A(\mathcal{P})$ or the volume $V(\mathcal{P})$  that we can ``associate with'' the event $\mathcal{P}$. Further, the total degrees of freedom, $\rho_g(\mathcal{P})\rho_g(\mathcal{Q})$, associated with two events $\mathcal{P}$ and $\mathcal{Q}$ is multiplicative
 while the primitive area/volume elements are additive. Hence this function $F$ should be an exponential.\footnote{This is an improvement over the discussion in my earlier works, like e.g., Ref.\cite{ijmpdreview2016} wherein I approximated it as a linear function. The lowest order results remain the same in either case but $\ln\rho_g(\mathcal{P})\propto A(\mathcal{P})$ makes better physical sense.} In terms of area, for example, we then get

\begin{equation}
 \ln 
 \left\{ 
 \begin{array}{c}
  \text{density of states of the}\\
  \text{quantum geometry at $\mathcal{P}$}
 \end{array}
 \right\}\quad
 \propto\quad
\left\{ 
 \begin{array}{c}
  \text{area ``associated with''}\\
  \text{the event $\mathcal{P}$}
 \end{array}
 \right\}
\end{equation}
That is,
$\ln\rho_g(\mathcal{P})\propto A(\mathcal{P})$.

We next need to give a precise meaning to the phrase, area (or volume) ``associated with'' the event $\mathcal{P}$. Any natural definition in standard Riemannian geometry will attribute zero volume and zero area to an event. But when we introduce the discreteness of the spacetime in terms of a zero-point length, we  find that \cite{paperD} the area associated with an event  becomes nonzero but the volume will still remain zero. What is more, this approach will introduce an arbitrary, constant norm, vector $n_a$ into the discussion. (Its norm is unity in the Euclidean sector and it will map to a  null vector with zero norm in the Lorentzian sector.). The area ``associated with'' an event will still be a fluctuating, indeterminate variable depending on a quantum degree of freedom $n^a$. I will show how all these results arise, concentrating on the area.

Let me first introduce the notion of an equi-geodesic surface,  
which can be done either in 
the Euclidean sector or in the Lorentzian sector;  I will work in the Euclidean sector. 
An equi-geodesic surface $\mathcal{S}$ is the set of all points at the same geodesic distance $\sigma$ from some specific point $P$, which we take to be the origin \cite{D1,D4,D5,D6}.  
I can now  ``associate'' an area element with a point $P$ in a fairly natural way  by the following limiting procedure:
(i) Construct an equi-geodesic surface $\mathcal{S}$ around a point $P$ at a geodesic distance $\sigma$. (ii) Calculate the area element $\sqrt{h}$ from the induced metric $h_{ab}$ on $\mathcal{S}$. (iii) Use the limit $\sigma\to0$ to define the area element associated with the point $P$.
 In standard differential geometry, we can show \cite{D8} that, in the limit of $\sigma \to 0$,  the area element, normalized to the flat spacetime value, is given by: 
\begin{align}
\frac{\sqrt{h}}{\sqrt{h_{flat}}}=\left(1-\frac{1}{6}\sigma ^{2}R^a_bn_an^b\right)
\label{gh}
\end{align}
where $n_a=\nabla_a\sigma$ is the normal to $\mathcal{S}$.  The second term involving  $\mathcal{E}\equiv R^a_bn_an^b$ gives the curvature correction to the area element of  an equi-geodesic surface. In fact, \eq{gh} describes a  textbook result in differential geometry and is often presented as a measure of the curvature at any event. 

As you can readily see from \eq{gh}, $\sqrt{h}\to0$  when $\sigma\to 0$ since, in this limit $\sqrt{h_{flat}}\propto\sigma^3$. Even though  the normal $n_a=\nabla_a\sigma$ becomes ill-defined in this limit, this ambiguity of the second term is irrelevant when $\sigma\to 0$ because of the $\sigma^2$ factor.
This is, of course,  to be expected. The existence of non-zero \md\ requires some kind of discrete structure in the spacetime. They cannot arise if  the spacetime is treated as a continuum all the way. (This is analogous to the fact that you can't associate a finite number  of molecules of a fluid with an event $P$ if the fluid is treated as a continuum all the way.)  Classical differential geometry, which leads to \eq{gh}, knows nothing about any discrete spacetime structure and hence cannot give you a nonzero $\rho_g$. To obtain a nonzero $\rho_g$, we need to know how the geodesic interval and the  metric get modified in the quantum description of spacetime. In particular, we would expect to have a $\sqrt{h}$  which does not vanish in the coincidence limit in  such a quantum description. I will now turn to the task of describing a spacetime metric which is modified by quantum gravitational effects from some general considerations.

There is a large amount of evidence (see e.g., \cite{D2a,D2b,D2c,D2d,D2e,D2f}) which suggests that a primary effect of quantum gravity will be to introduce into the spacetime  a zero-point length, by modifying the geodesic interval $\sigma^2(x,x')$ between any two events $x$ and $x'$  to a form like $\sigma^2 \to \sigma^2 + L_0^2$ where $L_0$ is a length scale of the order of the Planck length.\footnote{A more general modification will have the form $\sigma^2 \to S(\sigma^2)$  where the function $S(\sigma^2)$ satisfies the constraint $S(0) = L_0^2$ with $S'(0)$ finite. The results  described here   are  insensitive to the explicit functional form of  $S(\sigma^2)$. For the sake of  illustration, I will use the function $S(\sigma^2) = \sigma^2 + L_0^2$.} This  allows us to determine a quantum corrected, effective metric
along the following lines:
Just as the original $\sigma^2$ is obtained from the original metric $g_{ab}$, we demand that the geodesic interval $S(\sigma^2)$ --- which incorporates the effects of quantum gravity --- arises from a corresponding, quantum gravity-corrected metric \cite{D1}, which we will call the qmetric $q_{ab}$. Of course, no such local, non-singular $q_{ab}$ can exist; this is  because, for any such $q_{ab}$, the resulting geodesic interval will vanish in the coincidence limit,  by definition of the integral. The $q_{ab}(x,x')$ will  be a bitensor, which is singular at all events in the coincidence~limit $x\to x'$. The fact that the pair $(q_{ab},S(\sigma^2))$ should satisfy the same relationships as $(g_{ab},\sigma^2)$ is enough to
determine \cite{D4,D5,D7} the form of $q_{ab}$. We can then 
compute the area element ($\sqrt{h}\, d^3 x$) of an equi-geodesic surface 
using the  qmetric.  

The computation is 
straightforward  and --- for \mbox{$S(\sigma^2)=\sigma^2+L_0^2$} in $D=4$, though similar results \cite{paperD,D7} hold in the more general case in $D$ dimensions ---  we get:\footnote{You might think that the result in \eq{hfinal}  arises from the standard result in \eq{gh}, just by the simple replacement of $\sigma^2\to(\sigma^2+L_{0}^{2})$. This happens to be true but it turns out that this replacement trick does \textit{not} work for the volume element $\sqrt{q}$ which actually vanishes \cite{D4,D5,D7} when $\sigma\to0$. This is  why  each event has zero volume, but a finite area, associated with it! A further insight into this curious feature is provided by the following fact:
The leading order dependence of $\sqrt{q}d\sigma\approx\sigma d\sigma$ leads to the volumes scaling universally as $\sigma^2$ while the area measure is finite. This, in turn, leads to the result \cite{paperD} that \textit{the effective dimension of the quantum-corrected spacetime tends to $D=2$ close to Planck scales,} independent of the original $D$. Similar `dimensional reduction' has been noticed by several people \cite{z1}
in different, but specific, models of quantum gravity. The approach described here leads to this result in a fairly \textit{model-independent} manner.}
\begin{align}
\frac{\sqrt{h}}{\sqrt{h_{flat}}}
=\left[1-\frac{1}{6}(R_{ab}n^an^b)\left(\sigma ^{2}+L_{0}^{2}\right)\right];\qquad L_0^2=\frac{3L_P^2}{4\pi}
\label{hfinal}
\end{align}
(It will turn out that the correct Newtonian limit of the theory requires $L_0^2=3L_P^2/4\pi$ which is the value we will use.)
When $L_{0}^{2}\to0$, we recover the standard result in \eq{gh}, as we should. Our interest, however, is in the coincidence limit $\sigma^2\to0$ evaluated with finite $L_0$.
Something nice happens when we do this and we get a non-zero limit:
\begin{align}
\lim_{\sigma\to 0} \frac{\sqrt{h}}{\sqrt{h_{flat}}}
= \left[1-\frac{L_P^2}{8\pi}R_{ab}n^an^b\right]
\label{hlimit}
\end{align}
That is, the qmetric 
attributes to every point in the spacetime a finite area measure (but a zero volume measure)! 
So we define \cite{tpentropy} the dimensionless \dqg, as:\footnote{As you will see, the fact that the term involving $R_{ab}$  comes with a \textit{minus sign} in \eq{denast} is crucial for the success of our programme and we have \textit{no} control over it!} 
\begin{equation}
\ln\rho_g(x^i,n_a)\propto\lim_{\sigma\to 0}  \frac{\sqrt{h(x,\sigma)}}{\sqrt{h_{flat}(x,\sigma)}}
\propto \left[1-\frac{L_P^2}{8\pi}R_{ab}n^an^b\right]=\mu\left[1-\frac{L_P^2}{8\pi}R_{ab}n^an^b\right]
\label{denast}
\end{equation}
where $\mu$ is a dimensionless proportionality constant. As we mentioned earlier this is related to the choice of the measure;  alternatively, we can set $\mu=1$ in \eq{denast} and absorb it in the measure.

We see that $\rho_g$, defined by the first equality in \eq{denast}, depends on the extra, internal degree of freedom, $n_a$ which could take all possible values (at a given $x^i$) except for the constraint that it has  unit norm in the Euclidean space. \textit{This quantity is a relic of the discrete nature of the spacetime.} This is analogous to the $p_j$ which appears in the fluid distribution function $f(x^i,p_j)$ which can take all possible values at a given $x^i$ --- as a relic of the discrete nature of the fluid --- except for the constraint that it has a constant norm $p^2=m^2$. In the case of the fluid, we have $\langle p^\mu\rangle =P^\mu(x^i)$ where $P^\mu(x^i)$ is the mean momentum of the fluid in the continuum description and
 $\langle p^\mu p^\nu\rangle =P^\mu(x^i)P^\nu(x^i)+\Sigma^{\mu\nu}$, where the second term arises from the dispersion of momentum. Similarly  $n^a$ is a microscopic, fluctuating  variable such that its average over  fluctuations gives some vector field  $\langle n^a\rangle = \ell_E^a(x^i)$ of unit norm in the Euclidean continuum limit and $\langle n^in^j\rangle =\ell^i_E\ell^j_E+\sigma^{ij}$ where the second term $\sigma_{ij}$ represents higher corrections to the mean value.

So far, we have been working in the Euclidean sector with $n_a=\nabla_a\sigma$ being the unit normal to the equi-geodesic surface, $\sigma=$ constant. The limit $\sigma\to0$ in the Euclidean sector makes the equi-geodesic surface shrink  to the origin. \textit{But, in the Lorentzian sector, the same limit leads to the null surface which acts as the local Rindler horizon around the chosen event.} Therefore, in this limit, we need to identify $n^a$ and its mean value $\ell^a_E$ with Lorentzian null vectors. This is how an internal degree of freedom  enters into the \dqg, $\langle \ln\rho_g\rangle $, in the form of a null vector $\ell^a$.

\begin{figure}[t]
 \begin{center}
  \includegraphics[scale=0.48]{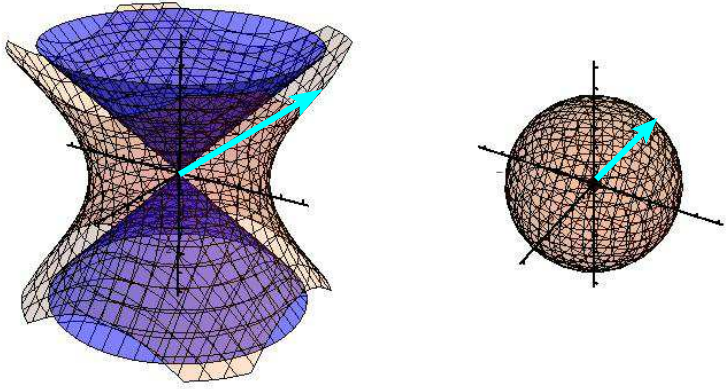}
 \end{center}
\caption{(a) Left: In the local inertial frame (in Lorentzian spacetime), the light cones originating from an event (taken to be the origin) are the null surfaces with $R^2-T^2 = 0$  with a normal $\ell_a$. The local Rindler observers on the hyperboloid $R^2-T^2 = \sigma^2 = $ constant located around these light cones  perceive a patch of the light cone as a local Rindler horizon with a non-zero temperature. The arrow denotes  the normal to the hyperbola. (b) Right: In the Euclidean sector, the hyperboloid $R^2-T^2 = \sigma^2$  becomes a sphere $R^2 + T_E^2 = \sigma_E^2$ and the  normal to the hyperboloid  becomes the normal to the sphere. The light cone $R^2 - T^2 =0$ translates to $R^2 + T_E^2 =0$, and hence collapses into the origin.  The limit $\sigma_E \to 0$, viz., approaching the origin in the Euclidean sector, corresponds to approaching the Rindler horizon in the Lorentzian sector. In this limit,  the hyperboloid becomes the two light cones emanating from $\mathcal{P}$. In the Euclidean sector, the direction of the normal to the sphere becomes ill-defined  when the radius of the sphere tends to zero. In the Lorentzian sector, it corresponds to   the normal to the null surface in the limit when the hyperboloid degenerates to the light cone.
 The dependence of $\rho_g$ on the normal $\ell^E_a$ to  equi-geodesic surface (in the Euclidean sector) is what translates into its dependence on the null normal $\ell_a$ in the Lorentzian sector.}
\label{fig:lightcones}
\end{figure}

To see this in some more detail,  consider the Euclidean version of the local Rindler frame. There are two natural ways of extending the null surface and the Rindler observer off the $TX$ plane in the Lorentzian sector. You can extend the null line (the 45 degree line in $TX$ plane) to the null \textit{plane}  $T=X$ in the spacetime. Alternatively, you can 
extend the null line  to the null \textit{cone} by $R^2-T^2=0$ where $R^2=X^2+Y^2+Z^2$ and the trajectories of Rindler observers to the hyperbolas in the  hyperboloid ($R^2-T^2=$ constant) which will go `around' the null cone in the Lorentzian spacetime (see the left part of Fig.~\ref{fig:lightcones}). Observers living on this hyperboloid will use their respective $X$ axes obtained by rotation. When we analytically continue to the Euclidean sector, the hyperboloid $R^2 - T^2 = \sigma^2$ will become a sphere $R^2 + T_E^2 = \sigma_E^2$ (see the right half of Fig.~\ref{fig:lightcones}). The light cone $R^2 - T^2 =0$, which transforms to $R^2 + T_E^2 =0$,  collapses into the origin. The local Rindler observers, living on the hyperboloid $R^2 - T^2 = \sigma^2$, will perceive local patches of the light cone $R^2 - T^2 =0$ as their local Rindler horizon (see the left half of Fig.~\ref{fig:lightcones}).
 Thus, taking the limit $\sigma_E \to 0$ in the Euclidean sector corresponds to approaching the local Rindler horizons in the Lorentzian sector with the hyperboloid degenerating into the light cones emanating from the event $\mathcal{P}$. The normal $\ell^E_a$ to the Euclidean sphere can be now identified with the normal to the local Rindler horizon $\ell_a$. The dependence of $\rho_g$ on $n_a$ in the Euclidean equi-geodesic surface  translates into its dependence on the null normal $\ell_a$ to which it maps in the Lorentzian sector.

We can summarize the net result of the exercise as follows: Take any event $x^i$ and a patch of any null surface ---  with normal $\ell_a(x)$ and passing through $x^i$ --- that can act as a horizon to the local Rindler observers. Identify $\langle n_a\rangle $ with $\ell_a(x)$ in the continuum limit. There are, of course, an infinite number of null patches (and normals) at any event. The quantum state of the microscopic geometry decides the expectation value $\langle n_a\rangle $ in a more fundamental description. Different quantum states will lead to  different null normals $\ell_a (x^i)$. Thus, the expectation value $\langle n_a\rangle $ actually corresponds to the \textit{set of} all null normals $\{\ell_a(x^i)\}$ at an event $x^i$ when we consider all possible quantum states. Then the \md\ associated with any event in the continuum spacetime is given by the mean value: 
\begin{equation}
\langle \ln\rho_g(x^i,n_a)\rangle =\mu\left[ 
 1-\frac{L_P^2}{8\pi}R_{ab}\ell^a\ell^b\right] +....
\label{denast1}
\end{equation}
where we have not displayed terms proportional to $R_{ab}\sigma^{ab}$ which are of higher order and independent of $\ell_a(x)$.

This approach brings to the center-stage the geodesic interval $\sigma^2(x,x')$ (rather than the metric) as the useful variable to describe spacetime geometry \cite{D7}. In the classical spacetime, both $\sigma^2(x,x')$ and $g_{ab}(x)$ contain the same amount of information and one is derivable from the other. But the geodesic interval $\sigma^2(x,x')$ is better suited to take into account quantum gravitational effects to a great extent.
Introducing a zero-point length in the spacetime by the modification $\sigma^2 \to S(\sigma^2)$ (with a finite $S(0)$) allows us to derive all the results presented above. 

This modification provides an operational principle for incorporating quantum gravitational effects at mesoscopic scales around any event $\mathcal{P}$. You can do this as follows: 
(a) Around the event $\mathcal{P}$, construct a local Euclidean version of the spacetime; this is almost always possible \textit{locally}. 
(b) Construct a synchronous frame around this event in which the line element has the form $ds^2 = d\sigma^2 + h_{\alpha\beta} dx^\alpha dx^\beta$. 
(c) Modify this metric to a quantum corrected qmetric, $q_{ab}(x,\mathcal{P})$  such that the geodesic interval acquires a zero-point length \cite{dawood1,paperD}. One can then explore the quantum corrected geometrical features using the bi-tensor $q_{ab}(x,\mathcal{P})$ in the neighbourhood of $\mathcal{P}$. In particular, this allows us to provide a natural definition of \dqg\ through \eq{denast}. 

 The procedure outlined above associates a constant norm, randomly fluctuating  vector field $n_i$ with every event in spacetime which is a relic from quantum gravity. This vector field can take arbitrary values throughout the spacetime and hence is described by some probability functional $P[n_i(x),x]$. While one cannot determine the form of $P[n_i(x),x]$ with our current knowledge of quantum gravity, we can use it to obtain the semi-classical limit of geometry with some reasonable assumptions.
   The $n_a$ gets mapped to the normal $\ell_a$ to the null surface  when we take the limit $\sigma \to 0$ in the Euclidean sector. This mapping, in turn, depends  on the fact that the condition $\sigma^2 (x,y) =0$ will lead to $x=y$ in the Euclidean space while it will be satisfied by all events connected by a null ray in the Lorentzian spacetime.

This expression in \eq{denast1} has a natural interpretation as the 
gravitational contribution to the heating rate (per unit area) of the null surface of which $\ell_a$ is the normal. Its integral over the null surface, $Q_g$, can be interpreted  as the gravitational contribution to the heat content of the null surface. These results arise because the term $R_{ab}\ell^a\ell^b$ is related to the concept of ``dissipation without dissipation'' \cite{sanvedtp} of the null surfaces. Let me explain briefly how this connection comes about. 

Construct the standard description of a null surface by introducing the complementary null vector $k^a$ (with $k^a\ell_a=-1$) and defining the 2-metric on the cross-section of the null surface by $q_{ab}=g_{ab}+\ell_ak_b+k_a\ell_b$. Define the expansion $\theta\equiv\nabla_a\ell^a$ and shear $\sigma_{ab}\equiv \theta_{ab}-(1/2)q_{ab}\theta$ of the null surface where $\theta_{ab}=q^i_aq^j_b\nabla_i\ell_j$. (It is convenient to take the null congruence  to be affinely parametrized.) One can then show that \cite{A19}:
\begin{equation}
-\frac{1}{8\pi L_P^2}R_{ab}\ell^a\ell^b\equiv\mathcal{D}+\frac{1}{8\pi L_P^2}\frac{1}{\sqrt{\gamma}}\frac{d}{d\lambda}(\sqrt{\gamma}\theta)
\label{rai}
\end{equation} 
where
\begin{equation}
 \mathcal{D}\equiv\left[2\eta \sigma_{ab}\sigma^{ab}+\zeta\theta^2\right]
\end{equation} 
is the standard expression for the viscous heat generation rate of a fluid with shear and bulk viscous coefficients \cite{A26,A27,membrane}
defined\footnote{The fact that the null fluid has negative bulk viscosity coefficient is well-known in the literature \cite{A26,A27,membrane}, especially in the context of the black hole membrane paradigm. So I will not pause to discuss this aspect.} as $\eta=1/16\pi L_P^2,\zeta=-1/16\pi L_P^2$. Ignoring the total divergence in \eq{rai}, we can identify the integral of $R_{ab}\ell^a\ell^b$ with: 
\begin{equation}
Q_{g}=-\frac{1}{8\pi L_P^2}\int \sqrt{\gamma}\, d^2x \, d\lambda\, R^a_b \ell_a\ell^b
=\int \sqrt{\gamma}\, d^2x \, d\lambda\, \mathcal{D}
=\int \sqrt{\gamma}\, d^2x \, d\lambda\, \left[2\eta \sigma_{ab}\sigma^{ab}+\zeta\theta^2\right]
\label{Qtoty}
\end{equation}
which represents the heat content of the null surface due to gravitational degrees of freedom and the integrand $\mathcal{D}$ is 
the rate of heating of the null surface due to gravitational degrees of freedom. 

Of course, we do not want the null surfaces to exhibit heating or dissipation! This is ensured by the presence of matter which is needed if $R_{ab}\neq0$. As we will see, the contribution to the heating from the microscopic degrees of freedom of the spacetime precisely cancels out the heating of  any null surface by the matter, on-shell. In fact, this allows us to reinterpret the field equation, expressed as 
\begin{equation}
-\frac{1}{8\pi L_P^2}R_{ab}\ell^a\ell^b +  T_{ab}\ell^a\ell^b=\mathcal{H}_g+\mathcal{H}_m= 0  
\label{zerodisp}
\end{equation} 
as a zero-dissipation principle. I will now introduce the physics of the matter sector and show how this comes about.

\section{Heat density of matter}
\label{sec:hdm}

I will next show how the combination $\mathcal{H}_m \equiv T_{ab} \ell^a \ell^b$ for any null normal $\ell_a$ can be thought of as the heat density  contributed by matter crossing  a null surface. This will eventually lead to the result in \eq{rhomresult}.

To acquire some preliminary insight, consider first the case of an ideal fluid, with $T^a_b = (\rho+p) u^au_b + p\delta^a_b$.  In this case, the combination $T^a_b \ell_a\ell^b$ is precisely the \textit{heat density} $\rho +p = Ts$ where $T$ is the temperature and $s$ is the entropy density of the fluid. (The last equality follows from the Gibbs-Duhem relation. We have chosen the null vector such that $(\ell.u)^2=1$ for simplicity.)  
But  $T^a_bu^bu_a$ is the energy density for \textit{any} kind of $T^a_b$, not just for that of an ideal fluid. How can we interpret $T^a_b \ell_a\ell^b$ as the \textit{heat} density in a  \textit{general} context when $T^a_b$ could describe any  source --- not necessarily a fluid --- for which concepts like temperature or  entropy do not exist intrinsically?  \textit{Remarkably enough,  you can do this!}. In any spacetime, around any event, we can introduce the  local Rindler observers who will indeed interpret $T^a_b \ell_a\ell^b$ as the heat density contributed by the matter to a null surface which they perceive as a horizon.  Let me describe this result: 

We begin by introducing a freely falling frame (FFF) with coordinates $(T, \mathbf{X})$ in a region around some fiducial event $\mathcal{P}$ in the spacetime.  Next, we transform from the FFF to a local Rindler frame (LRF) --- with coordinates $(t,\mathbf{x})$ --- using the transformations: $X=\sqrt{2ax}\cosh (at), T=\sqrt{2ax}\sinh (at)$ constructed using some acceleration $a$. (This transformation is valid for  $X>|T|$ and similar ones exist for other wedges.) One of the null surfaces passing though $\mathcal{P}$, will now get mapped to the $X=T$ surface of the FFF and will act as a patch of horizon to the local Rindler observers with the trajectories $x=$ constant.
This construction leads to a beautiful result \cite{du1,du2} in quantum field theory. The local vacuum state, appropriate for the freely-falling observers around $\mathcal{P}$, will appear to be a thermal state to the local Rindler observers, with a temperature proportional to their acceleration $a$: 
\begin{equation}
 k_BT = \left(\frac{\hbar}{c}\right) \left(\frac{a}{2\pi}\right)
\label{rindlertemp}
\end{equation} 
 (This acceleration $a$ can, of course, be related to other geometrical variables of the spacetime in different contexts). 
The existence of the  Davies--Unruh temperature tells us that around \textit{any} event, in \textit{any} spacetime, you will always have a class of observers who will perceive the spacetime to be hot.
 
Consider now the flow of energy associated with the matter, with some \textit{arbitrary} $T_{ab}$, that crosses the null surface. Nothing strange happens when this phenomenon is viewed in the FFF by the locally inertial observer. But the local Rindler observer, who attributes a temperature $T$ to the horizon,  sees it as a hot surface. Therefore, she will interpret the energy $\Delta E$, dumped on the horizon (by the matter that crosses the null surface in the FFF),  as  energy deposited on a \textit{hot} surface, thereby contributing a \textit{heat} content $\Delta Q=\Delta E$ to the surface. (Recall that, in the case of a \textit{black hole} horizon, an outside observer will find that any matter takes an infinite amount of time to cross the horizon,  thereby allowing for thermalization to take place. In a similar manner, a local Rindler observer will find that  matter takes a very long time to cross the local Rindler horizon.) It is easy to compute $\Delta E$ in terms of $T^a_b$. The LRF has an approximate Killing vector field, $\xi ^{a}$, generating the Lorentz boosts in the FFF, which coincides with a suitably defined\footnote{Since the null vectors have zero norm, there is an overall scaling ambiguity in any expression involving them. This can be resolved, in this case,  by considering a family of hyperboloids $\sigma^2\equiv X^2-T^2 =2ax =$ constant and treating the light cone as the (degenerate) limit $\sigma\to0$ of these hyperboloids. We set $\ell_a= \nabla_a\sigma^2\propto\nabla_a x$ and then take the corresponding limit.} null normal $\ell ^{a}$ of the null surface in the appropriate limit. Using the heat current that arises from the  energy current $T_{ab}\xi ^{b}$, we can compute  the total heat energy dumped on 
the null surface to be:
\begin{align}\label{Paper06_New_11}
 Q_{m}=\int \left(T_{ab}\xi ^{b}\right)d\Sigma ^{a}=\int T_{ab}\xi ^{b}\ell ^{a}\sqrt{\gamma}\ d^{2}x d\lambda
=\int T_{ab}\ell ^{b}\ell ^{a}\sqrt{\gamma}\ d^{2}x d\lambda
\end{align}
where we have used the fact that $\xi ^{a} \to \ell ^{a}$ on the null surface. 
Therefore, the combination
\begin{equation}
 \mathcal{H}_m\equiv \frac{dQ_{m}}{\sqrt{\gamma}d^{2}xd\lambda}=T_{ab} \ell^a\ell^b
\label{hmatter} 
\end{equation}
is indeed  the heat density (energy per unit area per unit affine time) of the null surface, contributed by matter crossing a local Rindler horizon. This interpretation is valid  for \textit{any} kind of $T^a_b$.

The entropy $S_m$ associated with the heat $Q_m$ in \eq{Paper06_New_11} is 
$S_m = Q_m/T_H$ where $T_H$ is a temperature introduced essentially for dimensional purposes. (A natural choice will be to take it to be the temperature associated with the acceleration of the Rindler observers very close to the horizon. But, as to be expected, none of the results will depend on its numerical value if we choose the measure of integration appropriately.) We have
\begin{equation}
S_m(\ell) = \frac{1}{T_H} \int d\lambda\, d^2 x \sqrt{\gamma}\, \mathcal{H}_m(x,\ell) 
= \mu \int \frac{d\lambda\, d^2 x \sqrt{\gamma}}{L_P^3} \, \left(L_P^4 \mathcal{H}_m(x,\ell)\right)
\label{tpnine}
\end{equation}  
where we have introduced suitable factors of $L_P$ to exhibit clearly the dimensionless nature of $S_m$ and defined $\mu \equiv (1/L_PT_H)$. 
Replacing the integration by a summation over the  spacetime events for conceptual clarity, we can write 
\begin{equation}
S_m(\ell) = \sum_x L_P^4 \mathcal{H}_m (x,\ell) \equiv \sum_x \ln \rho_m(x,\ell) = \ln \prod_x  \rho_m(x,\ell)
\label{tpten}
\end{equation} 
so that $\ln\rho_m(x,\ell)=L_P^4 \mathcal{H}_m (x,\ell)$ with a proportionality constant set to unity if we absorb the factor $\mu$ in the integration measure. (Alternatively, we could have set the proportionality constants  in \eq{rhogresult} and \eq{rhomresult} to be $\mu$ and used the measure without the factor $\mu$. The results will be the same.) 
This connects up with our discussion in Sec. \ref{sec:grfratoms}   and, in particular, with the result  in \eq{rhomresult} in the continuum limit. 
Note that the total number of degrees of freedom is correctly given by
\begin{equation}
\Omega_m(\ell) = \exp S_m(\ell) = \prod_x \rho_m = \prod_x \exp(L_P^4 \mathcal{H}_m ) = \exp\left[\mu \int \frac{d\lambda\, d^2 x \sqrt{\gamma}}{L_P^3} \, \left(L_P^4 \mathcal{H}_m\right)\right]
\label{tpeleven}
\end{equation} 
The first equality is the standard relation between the entropy and the degrees of freedom, the second expresses the result as a product over the degrees of freedom associated with each event and the third equality presents it in terms of the variable $\mathcal{H}_m = T^{ab}\ell_a\ell_b$.  

Since the matter stress tensor appears only through the combination $T^a_b \ell_a \ell^b$ in the theory, it is invariant  under the shift $T^a_b \to T^a_b + $ (constant)$\delta^a_b$ arising, for example, by the addition of a constant to matter Lagrangian.
 This is a symmetry of the matter sector, since the equations of motion for matter are unaffected by the addition of a constant to the matter Lagrangian. But this symmetry  is not respected by gravity in the conventional approaches, in which the metric is varied as a dynamical variable in an extremum principle. Any such approach is incapable of solving the cosmological constant problem because the shift $T^a_b \to T^a_b + $ (constant)$\delta^a_b$ will regenerate it. In the approach presented here, it is the heat density $\rho +p$ rather than the energy density $\rho$ which couples to gravity and \cc\ arises as an integration constant. As I will describe in Sec. \ref{sec:cc}, its value can also be determined within this approach in a self-consistent manner.

In the above discussion, we identified the average $\langle n_a\rangle =\ell_a(x)$ with the normal to the null surface
and hence the heat density of matter can be written as:
\begin{equation}
T^{ab}\ell_a\ell_b=T^{ab}\langle n_a\rangle \langle n_b\rangle \approx T^{ab}\langle n_an_b\rangle 
\end{equation} 
when we ignore higher order fluctuations. This suggests that the
expression $\langle \ln \rho_m\rangle $ again can arise  as the average value of 
\begin{equation}
\ln \rho_m \propto  L_P^4 T_{ab} n^a n^b                                          
\end{equation} 
which describes the (postulated) interaction of matter with 
 with the \md\ at these mesoscopic scales. The $x^i$ dependence of $\rho_m (x^i, n_a)$ arises through a symmetric divergence-free energy momentum tensor $T_{ab}(x)$. 
 
 We can also obtain the results of Sec.\ref{sec:grfratoms} directly using $n_a$ itself, along the following lines:
 The total \dqg\ when the internal degrees of freedom are described by a configuration $n_a(x)$ can  be written as:
 \begin{eqnarray}
 \Omega_{\rm tot}[n_a(x)] &=&\prod_{x}\, \rho_g (x^i, n_a)\, \rho_m (x^i, n_a) 
 =\exp \sum_x\left( \ln \rho_g + \ln \rho_m\right)\nonumber\\
 &=&\exp \mu\int dV_x\ L_P^4E^{ab}n_a(x) n_b(x)
 \label{omtot1}
\end{eqnarray}
with $E_{ab}\equiv T_{ab}-(1/\kappa)R_{ab}$.
The average over $n_a$ is given by:
\begin{equation}
\langle \Omega_{\rm tot}\rangle=\int\mathcal{D}n\ P[n_a(x)]\Omega_{\rm tot}[n_a(x)] 
=\int\mathcal{D}n\ P[n_a(x)]\exp \mu\int dV_x\ L_P^4E^{ab}n_a(x) n_b(x)
\end{equation} 
The  gravitational field equations arise on extremizing this expression for $\langle \Omega_{\rm tot}\rangle$.
The  $n^a$ dependent part of the exponential reduces to one proportional to 
\begin{equation}
\int dV_x E^a_b n_an^b\equiv \int dV_x \left(T^a_b (x) - \frac{1}{\kappa} R^a_b (x)\right) n_an^b;\qquad \kappa\equiv 8\pi L_P^2 
\label{tponen}
\end{equation} 
The extremum condition for this expression, with respect   to $n^a \to n^a + \delta n^a$, 
subject to the constraint $n^2=$ constant, leads to Einstein's equations, with a cosmological constant arising as an integration constant. 

There is a more interesting interpretation which is possible for the expression in \eq{tponen}. Recall that $T^{ab}\ell_a\ell_b$ has an unambiguous physical meaning as the heat density contributed by matter while $\mu=1/L_PT_H$ is the dimensionless Rindler horizon temperature. Therefore, one can think of $\exp-(1/T_H)(L_P^3 T^{ab}\ell_a\ell_b)$ as the Boltzmann factor for the energy in a Planck volume $L_P^3$ near the Rindler horizon. Of course, it doesn't make sense to talk about Planck volumes using $\ell_a$ but we can do it with $n_a$. This suggests the interpretation of
\begin{equation}
F \equiv \exp-\frac{1}{T_H}(L_P^3 E^a_b n_an^b)
=\exp-\frac{1}{T_H}\left(T^a_b (x) - \frac{1}{\kappa} R^a_b (x)\right) n_an^b
\label{boltz}
\end{equation} 
as the the Boltzmann factor for the excitation of the internal variable $n_a$. The extremum principle can be thought of as maximizing this expression. I hope to discuss this interpretation in detail in a later work.

\section{Cosmic information and the cosmological constant}\label{sec:cc}

A hallmark of any good paradigm is that it could throw light on issues which it was not originally designed to tackle. I will briefly explain how the approach presented above allows you to solve a major problem in theoretical physics, viz., the tiny value of cosmological constant. In fact, unlike any other approach to quantum gravity, this approach makes a falsifiable numerical prediction which, I believe, is vital for any formalism to be called science. I have described how this comes about in detail in previous works (see e.g., \cite{ijmpdreview2016,lwf}) and hence will be somewhat brief here. 

 Observations indicate that the evolution of our universe can be described by three different phases, viz., an inflationary phase very early on, followed by a radiation/matter dominated phase which lasted until recently, and an accelerated phase dominated by a small cosmological constant which has started in the near-past and will continue forever. These three phases can be characterized by three densities $\rho_{\rm inf}, \rho_{\rm eq}$ (which is the density of matter at the epoch when the matter and radiation densities were equal) and $\rho_\Lambda$. 
These three densities act as the signature of our universe, in the sense that the entire evolutionary history can be determined in terms of these numbers.
 Current observations determine $\rho_{\rm eq} $ and $\rho_\Lambda$ fairly accurately as: $\rho_{\rm eq} = [(0.86\pm 0.09) \ \text{eV} ]^4$ and 
 $\rho_\Lambda = [(2.26\pm 0.05)\times 10^{-3}  \text{eV}]^4$; we do not have a direct handle on $\rho_{\rm inf}$ but it is usually taken to be about $\rho_{\rm inf}\simeq (10^{15}\ \text{GeV})^4$.
So, in  standard cosmology, these three densities have absolutely no relation with each other and they are widely different. 

I  invite you to form a strange dimensionless number $I$ out of these three densities by the definition:
\begin{equation}
I= \frac{1}{9\pi} \, \ln \left( \frac{4}{27} \frac{\rho_{\rm inf}^{3/2}}{\rho_\Lambda\,\rho_{\rm eq}^{1/2}}\right) 
\label{strange1}
\end{equation} 
and evaluate its numerical value. Surprisingly enough, you will find that
\begin{equation}
I   \approx 4\pi \left( 1 \pm \mathcal{O} \left(10^{-3}\right)\right)
\label{strange2}
\end{equation} 
That is, $I = 4\pi$ to an accuracy of one part in thousand for the standard values expected for our universe. This should make you wonder why the  \eq{strange1} leads to such a pleasing value as $4\pi$, since it is not often that  such strange things happen.

The paradigm described here provides an answer \cite{hptpreview}.
The right hand side of \eq{strange1} can actually  be interpreted, in a well-defined manner, as the amount of of cosmic information accessible to an eternal observer \cite{hptpreview,ijmpdreview2016,lwf}; in fact it is this interpretation that leads to the rather strange combination in the first place. 
The reason this information content has a value $4\pi$ is related to the quantum microstructure of spacetime.\footnote{It is an \textit{observational fact}  that $I$, defined via \eq{strange1}, indeed has a numerical value $4\pi$ for our universe. It is hard to believe such a result is not telling us anything about our universe and can be ignored as ``just one of those things''!} 
This, in turn,  arises from the following facts: The effective dimension of the quantum-corrected spacetime becomes $D=2$ close to Planck scales, independent of the original $D$. This result can be derived, in a fairly model-independent manner using a renormalized quantum effective metric   \cite{paperD}. Similar results have been established earlier by several other authors (for a sample, see e.g., \cite{z1}) in a number of approaches to quantum gravity). This, in turn, implies that \cite{paperD,ijmpdreview2016} the unit of information associated with a quantum gravitational 2-sphere of radius $L_P$ is just $I_{\rm QG}=4\pi L_P^2/L_P^2=4\pi$. 

This relation allows us to determine $\rho_\Lambda$ in terms of $\rho_{\rm eq}$ and $\rho_{\rm inf}$ (both of which could eventually be determined from high energy physics): 
\begin{equation}
\rho_\Lambda = \frac{4}{27} \ \frac{\rho_{\rm inf}^{3/2}}{\rho_{\rm eq}^{1/2}} \ \exp \left( -36\, \pi^2\right)
\label{five}
\end{equation} 
If we take the typical values $\rho_{\rm inf} = (1.2 \times 10^{15}$ GeV)$^4, \rho_{\rm eq} = (0.86$ eV)$^4$, we get $\rho_\Lambda = (2.2 \times 10^{-3}$ eV)$^4$ which agrees very well with the observed value! 
That is, the idea that the cosmic information content accessible to an eternal observer, $I_c$, is equal to the basic quantum gravitational unit of information $I_{\rm QG} = 4\pi$, determines the numerical value of the \cc\ correctly.  

Theoretically,  the value of $\rho_{\Lambda}$ is the holy grail of cosmology. Observationally, however, we know  the values of $\rho_{\rm eq}$ and $\rho_\Lambda$ very well today but  have no direct handle on $\rho_{\rm inf}$. Using \eq{five}, we can actually \textit{predict} the value of $\rho_{\rm inf}$ in terms of the cosmologically determined parameters $\rho_{\rm eq}$ and $\rho_\Lambda$. We then find that  $\rho_{\rm inf}^{1/4} = 1.2 \times 10^{15}$ GeV. This is  the \textit{only} model with quantum gravitational inputs \textit{that leads to a falsifiable prediction}\footnote{In obtaining \eq{five}, I have assumed instantaneous reheating; ambiguities in the reheating dynamics can change this result by a factor of a few, leading to the prediction $\rho_{\rm inf}^{1/4} \approx (1-5) \times 10^{15}$ GeV.}.

More recently, we could improve this model in a significant manner and eliminate the inflationary phase altogether.\footnote{The summary below is based on results obtained in collaboration with H. Padmanabhan \cite{hptpnew}.} The new model is based on the idea  that the universe made a transition from a quantum pre-geometric phase to the classical Friedmann phase at an energy scale $E_{\rm QG} = \nu^{-1} E_{\rm Pl}$ where $\nu$ is a free parameter. We then make a single postulate: The total number of modes $N$ which enter the Hubble radius throughout the history of the universe is equal to $4\pi$. From this single assumption, one could obtain the following results: 
(a) The universe \textit{must have} a late-time accelerated phase. 
(b) The cosmological constant, which could drive this late-time acceleration, is given by:
\begin{equation}
\rho_\Lambda L_P^4 = \frac{4}{27} \  \left(\frac{3}{8\pi} \right)^{3/2}\frac{1}{\nu^6 (\rho_{\rm eq}L_P^4)^{1/2}} \ \exp \left( - 36\pi^2\right)
\label{fourpi}
\end{equation} 
(c) The primordial fluctuations (generated at the quantum to classical transitions) will be scale invariant and their amplitude is given by:
\begin{equation}
 \mathcal{A} = \left[\frac{k^3P(k)}{2\pi^2}\right]^{1/2}   = \frac{0.19 c_1} {\nu}
 \label{eqn6}
\end{equation} 
where $c_1$ is a numerical factor of order unity and whose exact value can be determined by more detailed analysis. 
Thus, a \textit{single} parameter $\nu$ determines  the values of \textit{both}\textit{}  $\rho_\Lambda$ as well as $\mathcal{A}$. There is absolutely no reason why we should be able to find a value for $\nu$ which gives the correct numerical values for \textit{both} these numbers. \textit{Incredibly enough, it turns out to be possible. }
Using the value of $\nu$ determined from \eq{fourpi}, we find that 
 $\mathcal{A}_{\rm theory} = 3.05 c_1 \times 10^{-5}$  which has to be compared with the observed value $\mathcal{A}_{\rm obs} \approx 4.69  \times 10^{-5}$. We see that the results are remarkably consistent with $c_1=1.54=\mathcal{O}(1)$.

These results highlight the notion of spacetime information and its role in gravitational dynamics, already seen in several other contexts \cite{ijmpdreview2016}. It also strengthens the viewpoint, suggested in Refs. \cite{ijmpdreview2016,lwf}, that the universe should \textit{not} be treated as a particular solution to the gravitational field equations but instead, be approached as a special dynamical system.

\section{Highlights and Discussion}
\label{sec:hds}

This approach, like any other approach to quantum gravity, involves both  algebraic features and conceptual features. The algebraic aspects are extremely simple\footnote{Unlike some others, I do not believe that the algebraic structure of quantum gravity has to be  complicated for it to be correct!} and, in fact, fairly trivial. 
Consider a functional of a null vector field $\ell_a$ given by \eq{av1} (or for $n_a$ given by \eq{tponen}). Demand that (a) this functional should be an extremum for the variations of the vector field which maintain its norm and (b)  the extremum condition should hold for all the null vector fields at any given event. It is trivial to show that these demands lead to Einstein's equations with the cosmological constant arising as an integration  constant.

What is really non-trivial about this approach are the conceptual aspects and the somewhat unfamiliar flow of logic. In particular, we need to understand the following aspects: (a) Where does the null vector field $n^a$ (or $\ell_a$) spring from? (b) What is the physical meaning of $R_{ab} n^a n^b$ which appears in the extremum principle? (c) What is the physical meaning of $T_{ab} n^a n^b$? (d) Why do we demand that the extremum condition should hold for all null vector fields at every event? 

I believe I have provided reasonably clear answers to all these questions. Just to summarize the previous discussion, the  answers --- in brief --- are as follows: (a) The vector field  $n^a$ arises as a relic of quantum gravity when we introduce a zero point length to the spacetime. It measures the \dqg\ at any given event through \eq{denast}. In fact, this is probably the most important new feature of this approach. (b) The physical meaning of  $R_{ab} n^a n^b$ is  provided by \eq{denast} when you think of $n_a$ as a fluctuating quantum variable. If we replace $n^a n^b$ by
$\langle n^a n^b\rangle \approx \ell^a\ell^b$, then this term acquires an interpretation as the ``dissipation without dissipation'' on the null surfaces through \eq{Qtoty}. (c) This is probably the most difficult issue. \textit{The Einstein equation $G_{ab}=\kappa T_{ab}$ equates apples to oranges.} The left hand side is geometrical while the right hand side, viz. matter, is made of discrete, quantized degrees of freedom bearing no relation to spacetime geometry. A description of matter in terms of $T_{ab}$ --- or even  in terms of the quantum expectation value $\langle T_{ab} \rangle$ --- is probably not going to be useful at the mesoscopic scales.\footnote{This is, in fact, a key issue not addressed satisfactorily in any approach to quantum gravity. We do not know how to describe matter fields close to Planck scales.}
We could couple both the  spacetime geometry and matter  to $n_a$ through the terms $R^a_b n_a n^b$ and $T^a_b n_a n^b$ respectively, thereby leading to an effective coupling between them.
 But this idea requires further investigation; one possible way forward could be through \eq{boltz}. But all these ideas still treat apples and oranges as though they are the same. We need a more unified way of generating both matter and geometry from some other fundamental building blocks, which is currently lacking. It is possible that the zero-dissipation-principle of \eq{zerodisp} tells us something important in this context but it is not clear what exactly is the message. So, I would consider the inclusion of matter into this formalism to be the major challenge at present. (d) This is easy. In this approach, 
the mean value of $n^a$ maps to a null vector field $\ell^a$. Different quantum  states of the geometry will lead to different $\ell^a$ at a given event. The demand that the extremum condition should hold for all $\ell^a$ is equivalent to demanding that the extremum condition holds for all quantum geometries; rather, the semi-classical spacetime arises only when this condition holds for all quantum geometries which are relevant for such a spacetime.

A real bonus, of course, is the fact that the approach provides a refreshingly different way of solving the \cc\ problem.

\section*{Acknowledgements}

I thank the organizers of DICE 2016 for excellent hospitality. Part of the discussion here is based on unpublished work \cite{hptpnew} done in collaboration with Hamsa Padmanabhan. I thank Sumanta Chakraborty, Sunu Engineer and Dawood Kothawala for useful discussions.

\end{document}